\documentclass[11pt]{article}
\usepackage{epsfig}

\author{A. Mondrag\'on\thanks{Invited talk presented at the XXI
    Symposium on Nuclear Physics. Oaxtepec, M\'exico, January 1998.}
   and E. Rodr\'{\i}guez-J\'auregui\\
  Instituto de F\'{\i}sica, UNAM, Apdo. Postal 20-364, 01000 M\'exico,
  D.F. M\'exico.}  \title{A parametrization of the CKM mixing matrix
  from a scheme of $ S(3)_L \bigotimes S(3)_R$ symmetry breaking}

\date{
}
\begin{document}
\maketitle

\begin{abstract}
  Recent interest in flavour or horizontal symmetry building (mass
  textures) has been spurred mainly by the large top mass and, hence,
  the strong hierarchy in quark masses.  Recently, various symmetry
  breaking schemes have been proposed based on the discrete,
  non-Abelian group $S(3)_{L}\otimes{S(3)_{R}}$ ,which is broken
  according to $S(3)_{L}\otimes{S(3)_{R}}\ \supset\ {S_{diag}}(3)\ 
  \supset\ {S_{diag}}(2)$ . The group $S(3)$ treats three objects
  symmetrically, while the hierarchical nature of the Yukawa matrices
  is a consequence of the representation structure, $\bf{1\oplus2}$,
  of $S(3)$, which treats the generations differently. Different
  ans\"{a}tze for the breaking of the sub-nuclear democracy give
  different Hermitian mass matrices, $M$, of the same modified
  Fritzsch type which differ in the numerical value of the ratio
  $M_{23}/M_{22}$.  A fit to the experimentally determined absolute
  values of the elements of the CKM matrix gives bounds on the
  possible values of the CP violating phase and gives a clear
  indication on the preferred symmetry breaking scheme. A
  parametrization of the CKM mixing matrix in terms of four quark mass
  ratios and one CP violating phase in very good agreement with the
  absolute value of the experimentally determined values of the CKM
  matrix elements is obtained.
\end{abstract}

PACS numbers: 12.15.Ft, 11.30.Er, 11.30.Hv, 12.15.Hh

\section{Introduction}

In this paper we try to express the entries in the $V_{CKM}$ mixing
matrix in terms of the ratios of the quark masses and one CP
violating phase.  Our approach is guided by the experimental
information on $V_{CKM}$ and a desire to have a simple pattern of
flavour symmetry breaking.

\section{Flavour permutational symmetry}

In this section, we review some previous work on the breaking of the
permutational flavour symmetry.

In the Standard Model, analogous fermions in different generations,
say u, c and t or d, s and b, have completely identical couplings to
all gauge bosons of the strong, weak and electromagnetic interactions.
Prior to the introduction of the Higgs boson and mass terms, the
Lagrangian is chiral and invariant with respect to any permutation of
the left and right quark fields. The introduction of a Higgs boson and
the Yukawa couplings give mass to the quarks and leptons when the
gauge symmetry is spontaneously broken. The quark mass term in the
Lagrangian, obtained by taking the vacuum expectation value of the
Higgs field in the quark Higgs coupling, gives rise to quark mass
matrices ${\bf M(d)}$ and ${\bf M(u)}$,
\begin{equation}\label{lag1}
{\cal L}_{Y} =\bf \bar{d}_{L}M(d)d_{R}+\bar{u}_{L}M(u)u_{R}+h.c.
\end{equation}

In this expression, $\bf d_{L,R}(X)$ and $\bf u_{L,R}(X)$ denote the
left and right quark $d$- and $u$-fields in the weak or coherent
basis.  A number of authors [1-8] have pointed out that realistic
quark mass matrices result from the flavour permutational symmetry
$S(3)_{L}\otimes S(3)_{R}$ and its spontaneous or explicit breaking.
The group $S(3)$ treats three objects symmetrically, while the
hierarchical nature of the mass matrices is a consequence of the
representation structure $\bf{1\oplus2}$ of $S(3)$, which treats the
generations differently. Under exact $S(3)_{L}\otimes S(3)_{R}$
symmetry the mass spectrum for either up or down quark sectors
consists of one massive particle in a singlet irreducible
representation and a pair of massless particles in a doublet
irreducible representation. To make explicit this assignment of
particles to irreducible representations of $S(3)_{L}\otimes
S(3)_{R}$, it will be convenient to make a change of basis from the
weak basis to a symmetry adapted or heavy basis through the unitary
transformation

\begin{equation}
\label{4}{\bf  M^{q}}_{H}={\bf U}^{\dagger}{{\bf M^{q}}_{W}}{\bf U},
\end{equation}
where
\begin{equation}
\label{6}{\bf U}={1\over \sqrt {6}}\pmatrix{
\sqrt {3} & 1 & \sqrt {2} \cr
-\sqrt {3} & 1 & \sqrt {2} \cr
0 & -2 & \sqrt {2} \cr
}.
\end{equation}
In the weak basis, the mass matrix with the exact $S(3)_{L}\otimes
S(3)_{R}$ symmetry reads
\begin{equation}\label{8}
{\bf M^{q}}_{3W}= {m_{3q}\over3}\pmatrix{
1 & 1 & 1 \cr
1 & 1 & 1 \cr
1 & 1 & 1 \cr
} _{W},
\end{equation}
where $m_{3q}$ denotes the mass of the third family quark, $t$ or $b$.  In
the symmetry adapted, or heavy basis, ${\bf M^{q}}_{3}$ takes the form
\begin{equation}\label{10}
{\bf M^{q}}_{3H}= m_{3q}\pmatrix{
0 & 0 & 0 \cr
0 & 0 & 0 \cr
0 & 0 & 1 \cr
}_{H},
\end{equation}
${\bf M^{q}}_{3H}$ is a singlet tensorial irreducible representation
of  $S(3)_{L}\otimes S(3)_{R}$. 

To generate masses for the second family, one has to break the
permutational symmetry $S(3)_{L}\otimes S(3)_{R}$ down to
$S(2)_{L}\otimes S(2)_{R}$. This may be done adding a term ${\bf
  M^{q}}_{2W}$ to ${\bf M^{q}}_{3W}$ such that
\begin{eqnarray}\label{12}
{\bf {M^{q}}}_{2W}={m_{3q}}\pmatrix{
\alpha & \alpha & \beta \cr
\alpha & \alpha & \beta \cr
\beta & \beta & \gamma \cr
}_{W}.
\end{eqnarray}

Still in the weak basis, the corresponding symmetry breaking term in
the Lagrangian is
\begin{eqnarray}\label{14}
{\bar{q}_{L}}({{\bf M^{q}}_{2W}}){q_{R}}&=
&\left\{ 2\alpha\left( { \bar{q}_{1L}+
\bar{q}_{2L}}\over\sqrt 2\right) \left( { q_{1R}+
q_{2R}}\over\sqrt 2\right)+\beta\sqrt 2\left[  \left( { \bar{q}_{1L}+
\bar{q}_{2L}}\over\sqrt 2\right)q_{3R}+\bar{q}_{3L}\left( { q_{1R}+
q_{2R}}\over\sqrt 2\right)\right]\right\}
\nonumber\\&+&{\gamma }\bar{q}_{3L}q_{3R}+h.c.
\end{eqnarray}

Notice that the symmetry breaking term depends only on the fields
$({q_{1}(X)+q_{2}(X)})/\sqrt 2$ and $q_{3}(X)$. Thus, the symmetry
breaking pattern is defined by requiring a well defined behaviour of
$\bar{q}_{L}({{\bf M^{q}}_{2W}})q_{R}$ under the exchange of the
fields ${(q_{1}(X)+q_{2}(X)})/\sqrt 2$ and $q_{3}(X)$.  There are only
two possibilities, either $\bar{q}_{L}({\bf M^{q}}_{2W})q_{R}$ is
symmetric or antisymmetric under the exchange of
$({q_{1}(X)+q_{2}(X)})/\sqrt 2~$ and $q_{3}(X)$.

In the antisymmetric breaking pattern \cite{9}, ${\bf M^{q}}_{2W}$
takes the form
\begin{eqnarray}\label{16}
{\bf {M^{q}}}_{2A,W}={m_{3q}}\pmatrix{
\alpha & \alpha & 0 \cr
\alpha & \alpha & 0 \cr
0 & 0 & -2\alpha \cr
}_{W} ,
\end{eqnarray}
in the heavy basis ${\bf {M^{q}}}_{2A}$ takes the form 
\begin{eqnarray}\label{18}
{\bf{\it M^{q}}}_{2A,H}={2m_{3q}\over 3}\pmatrix{
0 & 0 & 0 \cr
0 & -\alpha & {2\sqrt 2}\alpha \cr
0 & {2\sqrt 2}\alpha & \alpha \cr
}_{H} .
\end{eqnarray}
${\bf {M^{q}}}_{2A}$ has only one free parameter, $\alpha$, which is a
measure of the amount of mixing of the singlet and doublet irreducible
representations of $S(3)_{L}\otimes S(3)_{R}$.

In the symmetric breaking pattern \cite{1}, ${\bf {M^{q}}}_{2}$ is
given by

\begin{eqnarray}\label{20}
{\bf M^{q}}_{2S,W}={m_{3q}}\pmatrix{
\alpha & \alpha & \beta \cr
\alpha & \alpha & \beta \cr
\beta & \beta & 2\alpha \cr
}_{W} ,
\end{eqnarray}
and in the heavy basis ${\bf M^{q}}_{2S}$ becomes
\begin{eqnarray}\label{22}
{\bf{\it M^{q}}}_{2S,H}={2m_{3q}\over 3}\pmatrix{
0 & 0 & 0 \cr
0 & 2(3\alpha-2\beta) &-{\sqrt 2}\beta \cr
0 & -{\sqrt 2}\beta & 2(3\alpha+2\beta) \cr
}_{H} .
\end{eqnarray}
In this case ${\bf M^{q}}_{2S}$ has two free parameters, $\alpha$,
which is a shift in the masses of the second and third families, and
$\beta$, which produces a mixing of the singlet and doublet irreducible
representations.

In the heavy basis the symmetry breaking pattern is
usually characterized in terms of $\beta =-{ ({3\sqrt {2}})\over
  {4m_{3q}}}{({{\bf M^{q}}_{2S,H}})_{23}}~$ and the ratio
\begin{eqnarray}\label{24}
{(Z_{q}})^{1/2}={{\left( {\bf{\it{M^{q}}}}_{2S,H}\right)_{23}}\over 
{\left ( {{\bf{\it{M^{q}}}}_{2S,H}}\right )_{22}}}~.
\end{eqnarray}

In the antisymmetric breaking pattern there is no shift of the masses
and $Z_{q}$ takes the value 8. In the symmetric pattern, $Z_{q}$ is a
continuous parameter. Different values for $Z_{q}$ have been proposed
in the literature by various authors. If no shifting is allowed
\cite{1}, $\alpha=0$, and $Z_{q}=1/8$; Fritzsch and Holtsmanp\"{o}tter
\cite{7} put $\alpha = \beta$ which gives $Z_{q}=1/2$; Z. Z. Xing
\cite{10} takes $Z_{q}=2$ which is equivalent to setting $\alpha =
{1\over 2}\beta$. In the absence of a symmetry motivated argument to
fix the value of $Z_{q}$, other choices are possible. In this paper,
we will look for the best value of $Z_{q}$ by comparison with the
experimental data on the CKM mixing matrix and the Jarlskog invariant.

In order to give mass to the first family, we add another term
${\bf{M^{q}}}_{1}$ to the mass matrix. It will be assumed that
${\bf{M^{q}}}_{1}$ transforms as the mixed symmetry term of the
doublet complex tensorial representation of the $S(3)_{diag}$ diagonal
subgroup of $S(3)_{L}\otimes S(3)_{R}$. Putting the first family in a
complex representation allows us to have a CP violating phase. Then,
in the weak basis, ${\bf{M^{q}}}_{1}$ is given by
\begin{eqnarray}\label{26}
{\bf{M^{q}}}_{1W}={m_{3}\over {\sqrt 3}}\pmatrix{
A_{1} & iA_{2} & -A_{1}-iA_{2} \cr
-iA_{2} & -A_{1} & A_{1}+iA_{2} \cr
-A_{1}+iA_{2} & A_1-iA_{2} & 0 \cr
} .
\end{eqnarray}
In the symmetry adapted or heavy basis, ${\bf{M^{q}}}_{1}$ takes the
form
\begin{eqnarray}\label{28}
{\bf{M^{q}}}_{1H}={m_{3}} \pmatrix{ 
0 & {A_{q}}{e}^{-i{\phi _{q}}} & 0 \cr
{A_{q}}{ e}^{+i{\phi _{q}}} & 0 & 0 \cr
0 & 0 & 0 \cr
} .
\end{eqnarray}
Finally, adding the three mass terms, we get the mass matrix ${\bf
  M^{q}}$.

\section{Modified Fritzsch texture}

In the heavy basis, ${\bf{M^{q}}}_{H}$ has a modified Fritzsch texture
of the form
\begin{equation}\label{30}
{{\bf{\it M}}^{q}}_{H}={m_{3q}}\pmatrix{
0 & {A_{q}}{{e}}^{-i{\phi_{q}}} & 0 \cr
{A_{q}}{{e}}^{+i{\phi_{q}}}& D_{q} & B_{q} \cr
0 & B_{q} & C_{q} \cr
} .
\end{equation}
{}From the strong hierarchy in the quark masses,
$m_{3q}>>m_{2q}>>m_{1q}$, we expect $C_{q}$ to be very close to unity.
Therefore, it will be convenient to introduce a small parameter
$\delta _{q}$ through the expression
\begin{equation}\label{32}
{C}_{q} \equiv 1 - \delta_{q}
\end{equation}
The other entries in the mass matrix, namely $A_{q}$, $B_{q}$ and
$D_{q}$, may readily be expressed in terms of the mass ratios
\begin{eqnarray}\label{34}
\tilde{m}_{1q}={m_{1q}\over m_{3q}}\qquad\qquad {\rm ~and~} 
\qquad\qquad\tilde{m}_{2q}={m_{2q}\over m_{3q}}
\end{eqnarray}
and the parameter $\delta_{q}$. Computing the invariants of ${{\bf
    M}^{q}}_{H}$, $tr~({{\bf M}^{q}}_{H})$, $tr~\left ( {{\bf
      M}^{q}}_{H}\right ) ^{2}$ and $det~({{\bf M}^{q}}_{H})$ from
(\ref{30}) and comparing with the corresponding expressions in terms
of the mass eigenvalues $(m_{1},-m_{2},m_{3})$, we get
\begin{eqnarray}\label{36}
{{A}_{q}}^{2}={\tilde{m}_{1q}\tilde{m}_{2q} \over{1 - \delta_{q}}} ,
\end{eqnarray}
\begin{eqnarray}\label{37}
{{B}_{q}}^{2}=\delta_{q}\left( (1-\tilde{m}_{1q}+
\tilde{m}_{2q}-\delta_{q})  -{\tilde{m}_{1q}\tilde{m}_{2q} 
\over{1 - \delta_{q}}} \right),
\end{eqnarray}
and
\begin{eqnarray}\label{38}
{D}_{q}=\tilde{m}_{1q}-\tilde{m}_{2q}+\delta_{q}  ~.
\end{eqnarray}

If each possible symmetry breaking pattern is now characterized by the
parameter $Z_{q}$,
\begin{eqnarray}\label{40}
{{Z}_{q}}^{1/2}={{B}_{q}\over {D}_{q} } ~,
\end{eqnarray}
we obtain the following cubic equation for $\delta_{q}$
\begin{eqnarray}\label{42}
\delta_{q}\left\{(1+\tilde{m}_{2q}-\tilde{m}_{1q}-\delta_{q}) 
(1 - \delta_{q})-\tilde{m}_{1q}\tilde{m}_{2q}\right\}-{Z}_{q}(1-\delta_q)
(-\tilde{m}_{2q}+\tilde{m}_{1q}+\delta_{q})^{2}=0 ~.
\end{eqnarray}

The small parameter $\delta_{q}$ in eqs.~(\ref{32}) and (\ref{37}) is
the solution of  (\ref{42}) which vanishes when $Z_{q}$
vanishes. Then, the vanishing of $Z_{q}$ implies that $B_{q}=0$ and
$C_{q}=1$, or equivalently, there is no mixing of singlet and doublet
irreducible representations of   $S(3)_{L}\otimes S(3)_{R}$,  and the
heaviest quark in each sector is in a pure singlet representation.

\begin{figure}[htbp]
  \begin{center}
    \input{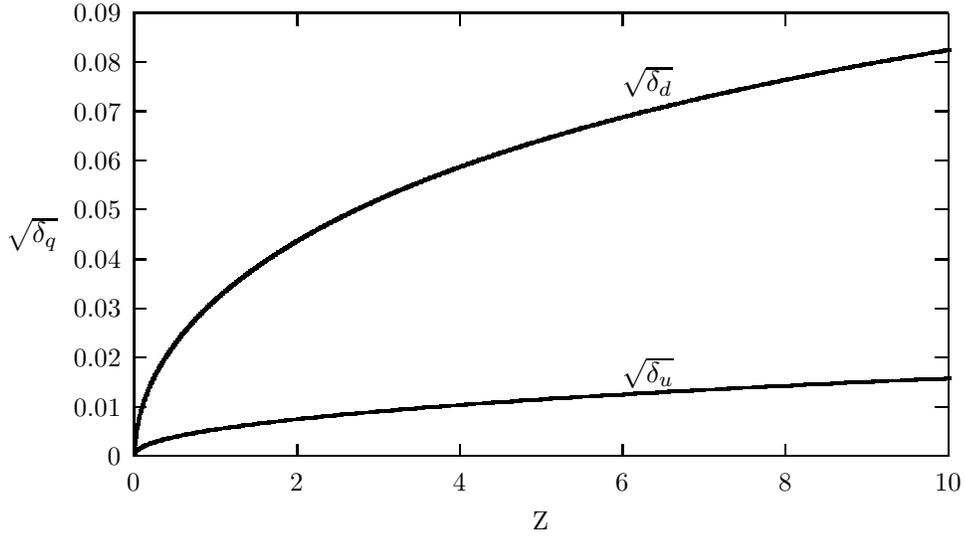}

    \caption{The square root of the parameters $\delta_{u}$, $\delta_{d}$
      is shown as function of the ratio $Z_{q}$. The value $Z\approx
      5/2$ which satisfies the constraining condition (\ref{64}) may
      be read from the figure.}
    \label{Figure 1}
  \end{center}
\end{figure}

In fig.~1, $\sqrt {\delta_{q}}$ is shown as function of $Z_{q}$. It
may be seen that, as $Z_{q}$ increases, $\sqrt {\delta_{q}(Z_{q})}$
increases with decreasing curvature. For very large values of $Z_{q}$,
$\sqrt {\delta_{q}(Z_{q})}~$ goes to the asymptotic limit
$\delta_{q}^{1/2}(\infty)=(\tilde{m}_{2q}-\tilde{m}_{1q})Z_q^{1/2}$.  Hence,
$\delta_{q}(Z_{q})$ is a small parameter, $\delta_{q}(Z_{q})<<1$, for
all values of $Z_{q}$. For large values of $Z_{q}$, say $Z_{q}\geq
20$, $\delta_{q}(Z_{q})$ is not sensitive to small changes in $Z_{q}$.

{}From eq.~(\ref{42}), we derive an approximate solution for
$\delta_{q}(Z_{q})$ valid for small values of $Z_{q}$ ($Z_{q}\leq
10$). Computing in the leading order of magnitude
\begin{eqnarray}\label{44}
 \delta_{q}\left( Z_{q} \right)\approx {Z_{q}
\left(   \tilde{m}_{2q}-\tilde{m}_{1q} \right)^{2}
\over \left(1-\tilde{m}_{1q} \right)\left( 1  +
\tilde{m}_{2q} \right)+2Z_{q}\left(   \tilde{m}_{2q}-\tilde{m}_{1q} 
\right)(1+{1\over 2}(\tilde{m}_{2q}-\tilde{m}_{1q}))}~.
\end{eqnarray}

\section{The CKM mixing matrix}

The Hermitian mass matrix ${\bf M_{q}}$ may be written in terms of a
real symmetric matrix ${\bf \bar{M}_{q}}$ and a diagonal matrix of
phases ${P_{q}}$ as follows
\begin{eqnarray}\label{46}
{\bf M^{(q)}}={{P}_{q}}{\bf \bar {M}^{(q)}}{{P}_{q}}^{\dagger}.
\end{eqnarray}
The real symmetric matrix ${\bf \bar {M}_{q}}$ may be a brought to
diagonal form by means of an orthogonal transformation
\begin{eqnarray}\label{48}
{\bf \bar {M}^{(q)}}={{O}_{q}}{\bf M^{(q)}_{diag}}{{O}_{q}}^{T},
\end{eqnarray}
where
\begin{eqnarray}\label{50}
{\bf {M^{(q)}}_{diag}}=m_{3q}~diag~[\tilde{m}_{1},-\tilde{m}_{2},1]
\end{eqnarray}
with subscripts 1,2,3 referring to $u,c,t$ in the $u$-type sector and
$d,s,b$ in the $d$-sector.  After the diagonalization of the mass
matrices ${\bf M^{(q)}}$, one obtains the CKM mixing matrix as
\begin{eqnarray}\label{52}
V_{CKM}={{O_{u}}^{T}}P^{(u-d)}{O}_{d},
\end{eqnarray}
where $P^{(u-d)}$ is the diagonal matrix of the relative phases.  

In the heavy basis, where $\bf {M^{(q)}}$ is given by (\ref{30})
-(\ref{42}), $P^{(u-d)}$ is
\begin{eqnarray}\label{54}
P^{(u-d)}=diag~[e^{-i\alpha},1,1],
\end{eqnarray}
$\alpha = \phi_u - \phi_d$, and the orthogonal matrix ${O_{q}}$ is
given by \cite{11}, \cite{12},
\begin{equation}\label{56}
{{O}}^{q}=\pmatrix{
\left(\tilde{m}_{2}{\rm {f}}_{1}/{\Delta}_{1} \right)^{1/2} & 
-\left(\tilde{m}_{1}{\rm {f}}_{2}/{\Delta}_{2} \right)^{1/2} & 
\left(\tilde{m}_{1}\tilde{m}_{2}{\rm {f}}_{3}/{\Delta}_{3} 
\right)^{1/2} \cr
\left({C}^{q}\tilde{m}_{1}{\rm {f}}_{1}/{\Delta}_{1} 
\right)^{1/2} & \left({C}^{q}\tilde{m}_{2}{\rm {f}}_{2}/{\Delta}_{2} 
\right)^{1/2}  & \left({C}^{q}{\rm {f}}_{3}/{\Delta}_{3} 
\right)^{1/2}  \cr
-\left(\tilde{m}_{1}{\rm {f}}_{2}{\rm {f}}_{3}/{\Delta}_{1} 
\right)^{1/2} & -\left(\tilde{m}_{2}{\rm {f}}_{1}
{\rm {f}}_{3}/{\Delta}_{2} \right)^{1/2} & \left({\rm {f}}_{1}
{\rm {f}}_{2}/{\Delta}_{3} \right)^{1/2} \cr
}
\end{equation}
where
\begin{eqnarray}\label{58}
{\rm {f}}_{1}=1-\tilde{m}_{1q}-{\delta}_{q}\qquad\qquad 
{\rm {f}}_{2}=1+\tilde{m}_{2q}-{\delta}_{q}\qquad\qquad
{\rm {f}}_{3}={\delta}_{q} \qquad\qquad\qquad\qquad\\
{\Delta}_{1}=(1-\delta_q)\left( 1-\tilde{m}_{1q} \right)\left( 
\tilde{m}_{2q}+\tilde{m}_{1q} \right)\qquad\qquad
{\Delta}_{2}=(1-\delta_q)\left( 1+\tilde{m}_{2q} \right)\left( 
\tilde{m}_{2q}+\tilde{m}_{1q} \right)\qquad\qquad\nonumber\\
{\Delta}_{3}=(1-\delta_q)\left( 1+\tilde{m}_{2q} \right)\left( 
1-\tilde{m}_{1q} \right)\qquad\qquad\qquad\qquad\qquad
\end{eqnarray}

{}From eqs.~(\ref{46}) - (\ref{58}), all entries in the $V_{CKM}$
matrix may be written in terms of four mass ratios: $(\tilde{m}_{u},
\tilde{m}_{c}, \tilde{m}_{d}, \tilde{m}_{s})$ and three free real
parameters : $\delta_{u}, \delta_{d}$ and $\alpha=\phi_{u}-\phi_{d}$.
The phase $\alpha$ measures the mismatch in the $\bar{S}_{diag}(2)$
symmetry breaking in the $u$- and $d$-sectors. It is this phase, and
consequently, that mismatch, which is responsible for the weak
violation of CP.

\section{The best value of $Z_{q}$}

At this stage in our argument, a question comes naturally to mind.
Does a comparison of the theoretical mixing matrix ${V_{CKM}}^{th}$
with the experimentally determined ${V_{CKM}}^{exp}$ give any clue
about the actual pattern of $S(3)_{L}\otimes S(3)_{R}$ symmetry
breaking realized in nature? or phrased differently: What are the best
values for $Z_{u}$ and $Z_{d}$? What is the best value for $\alpha$? Do
these values correspond to some well defined symmetry breaking
pattern?  

As a preliminary step in the direction of finding an answer
to these questions, we made a $\chi^{2}$ fit of the exact expressions
for the absolute value of the entries in the mixing matrix, that is
$|{V_{CKM}}^{th}|$ and the Jarlskog invariant $J^{th}$ to the
experimentally determined values of $|{V_{CKM}}^{exp}|$ and $J^{exp}$.
We left the mass ratios fixed at the values \cite{13}
\begin{eqnarray}\label{60}
{\tilde{m}}_{u}=0.00002\qquad\qquad{\tilde{m}}_{c}=0.00517
\end{eqnarray}
\begin{eqnarray}\label{62}
{\tilde{m}}_{d}=0.0019\qquad\qquad{\tilde{m}}_{s}=0.035~,
\end{eqnarray}
and we looked for the best values of the three parameters $\delta_{u},
\delta_{d}$ and $\alpha$.  We found the following results:

I.- Excellent fits of similar quality, $\chi^{2}\leq 0.3$, were
  obtained for a continuous family of values of the parameters
  $(\delta_{u},\delta_{d})$.

II.- In each good quality fit, the best value of $\alpha$ was
  fixed without ambiguity.

III.- The best value of $\alpha$ was nearly stable against large
changes in the values of $(\delta_{u},\delta_{d})$ which produced fits
of the same good quality.

IV.- In all good quality fits, the difference
$\sqrt{\delta_{d}}-\sqrt{\delta_{u}}$ takes the same value
\begin{eqnarray}\label{64}
\sqrt{\delta_{d}}-\sqrt{\delta_{u}}\simeq 0.040
\end{eqnarray}

These results may be understood if we notice that not all entries in
${V_{CKM}}^{th}$ are equally sensitive to variations of the different
parameters. Some entries, like $V_{us}$ are very sensitive to changes
in $\alpha$ but are almost insensitive to changes in
$(\delta_{u},\delta_{d})$ while some others, like $V_{cb}$ are almost
insensitive to changes in $\alpha$ but depend critically on the
parameters $\delta_{u}$ and $\delta_{d}$.

{}From eqs.~(\ref{52}), (\ref{56}) and (\ref{58}) we obtain
\begin{eqnarray}\label{66}
&V_{us}&=-\left( {{\tilde{m}}_{c}{\tilde{m}}_{d}\over 
{\left( 1-{\tilde{m}}_{u} \right)\left( {\tilde{m}}_{c}+
{\tilde{m}}_{u} \right)\left( 1+{\tilde{m}}_{s}\right)
\left(  {\tilde{m}}_{s}+{\tilde{m}}_{d}\right)}}\right)^{1/2} 
\left({\left(1-{\tilde{m}}_{u}-{\delta}_{u}  \right) \left(1+
{\tilde{m}}_{s}-{\delta}_{d}\right)\over {\left( 1-{\delta}_{u} 
\right) 
\left(1-\tilde {\delta}_{d} \right)}}\right)^{1/2}
e^{i\alpha}\cr &+&\left( {{\tilde{m}}_{u}{\tilde{m}}_{s}
\over {\left( 1-{\tilde{m}}_{u} \right)\left( {\tilde{m}}_{c}+
{\tilde{m}}_{u} \right)\left(  {\tilde{m}}_{d}+{\tilde{m}}_{s}
\right)}} \right) ^{1/2} \cr &\times&\left\{{ \left( 1-
{\tilde{m}}_{u}-\delta_{u}\right)^{1/2}\left( 1+{\tilde{m}}_{s} 
-{\delta}_{d}\right)^{1/2}\over{({1+{\tilde{m}}_{d}})^{1/2}( 1
-{\delta_{d}})^{1/2}}}+\left( (1+{\tilde{m}}_{c}-{\delta}_{u})
\delta_{u}\over {1-\delta_{u}}\right)^{1/2}\left((1-{\tilde{m}}_{d}
-{\delta}_{d})\delta_{d}\over {1-\delta_{d}}\right)^{1/2} \right\}
\end{eqnarray}
In the leading order of magnitude,
\begin{eqnarray}\label{68}
\mid V_{us} \mid \approx \mid\sqrt{{\tilde m}_{d} / {\tilde m}_{s}} 
e^{i\alpha} - \sqrt{{\tilde m}_{u} / {\tilde m}_{c}} 
\mid\left( 1- {\tilde m}_{u} / {\tilde m}_{c}-{\tilde m}_{d} / 
{\tilde m}_{s}\right)^{1/2}
\end{eqnarray}
Hence,
\begin{eqnarray}\label{70}
\cos \alpha ~\approx {{{\tilde m}_{d} / {\tilde m}_{s}+
{\tilde m}_{u} / {\tilde m}_{c}-\mid V_{us} \mid^{2}\left(  1- 
{\tilde m}_{u} / {\tilde m}_{c}-{\tilde m}_{d} / {\tilde m}_{s}
\right)}\over {2 \sqrt{({\tilde m}_{d} / {\tilde m}_{s})
({\tilde m}_{u} / {\tilde m}_{c})} }}
\end{eqnarray}
Substitution of $|{V_{us}}^{exp}|^{2}$ for $|V_{us}|^{2}$ and the
numerical value of the mass ratios, obtained from (\ref{60}) and
(\ref{64}), in (\ref {70}) gives
\begin{eqnarray}\label{72}
72^{\circ}\leq\alpha\leq82^{\circ}
\end{eqnarray}
with a mean value 
\begin{eqnarray}\label{74}
\bar{\alpha}=77^{\circ}~,
\end{eqnarray}
in good agreement with the best values extracted from the preliminary
$\chi^{2}$ fit.

Similarly, ${V_{cb}}^{th}$ is given by
\begin{eqnarray}\label{76}
{V_{cb}}^{th}&=&-\left\{ \tilde{m}_{u}\left( 1+\tilde{m}_{c}
-\delta_{u} \right)\over{\left( 1-\delta_{u} \right)\left( 1+
\tilde{m}_{c} \right)\left( \tilde{m}_{u}+\tilde{m}_{c} \right)} 
\right\}^{1/2}\left\{ \left(  \tilde{m}_{d}\tilde{m}_{s}
\delta_{d}\right) \over{\left(  1-\delta_{d}\right)\left( 1+
\tilde{m}_{s} \right) \left( 1-\tilde{m}_{d} \right)} \right\}^{1/2}
e^{i\alpha}\cr&+&\left\{ \tilde{m}_{c}\left( 1+\tilde{m}_{c}-
\delta_{u} \right)\over{\left( 1-\delta_{u} \right)\left( 1+
\tilde{m}_{c} \right)\left( \tilde{m}_{u}+\tilde{m}_{c} \right)} 
\right\}^{1/2}\left\{ \delta_{d}\over{\left( 1+\tilde{m}_{s} 
\right) \left( 1-\tilde{m}_{u} \right)} \right\}^{1/2}
\cr&-&\left\{ \tilde{m}_{c}\delta_{u}\left( 1-\tilde{m}_{u}-
\delta_{u} \right)\over{\left( 1-\delta_{u} \right)\left( 1+
\tilde{m}_{c} \right)\left( \tilde{m}_{u}+\tilde{m}_{c} \right)} 
\right\}^{1/2}\left\{ \left( 1- \tilde{m}_{d}-\delta_{d}\right)
\left( 1+\tilde{m}_{s}-\delta_{d} \right) \over{\left(  1-
\delta_{d}\right)\left( 1+\tilde{m}_{s} \right) \left( 1-
\tilde{m}_{d} \right)} \right\}^{1/2}.
\end{eqnarray}
Therefore, in the leading order of magnitude, $|V_{cb}|$ is
independent of $\alpha$ and given by
\begin{eqnarray}\label{78}
\mid {V}_{cb} \mid\approx\sqrt{\delta_{d}} -\sqrt{\delta_{u}}~.
\end{eqnarray}
Hence, good agreement with $|{V_{cb}}^{exp}|\approx 0.039$ \cite{13}
requires that
\begin{eqnarray}\label{80}
\sqrt{\delta_{d}} -\sqrt{\delta_{u}}\approx 0.039
\end{eqnarray}
at least for one pair of values $(\delta_{u},\delta_{d})$.  As
explained above, in the preliminary $\chi^{2}$ fit to the data, it was
found that (\ref{80}) is satisfied almost exactly, not just for one
pair of values of $\delta_{u}$ and $\delta_{d}$, but for a continuous
range of values of these parameters in which $\delta_{u}$ and
$\delta_{d}$ change by more than one order of magnitude.  

Therefore, eq.~(\ref{64}) may be used as a constraining condition
on the possible values of $(\delta_{u},\delta_{d})$.  In this way, we
eliminate one free parameter in ${V_{CKM}}^{th}$ without spoiling the
good quality of the fit. 

If instead of taking $(\delta_{u},\delta_{d})$ as free parameters, we
use the parameters $(Z_{u},Z_{d})$ to characterize the pattern of
symmetry breaking, we should write $\delta_{q}$ as function of $Z_{q}$
in (\ref{64}) to restrict the possible values of $(Z_{u},Z_{d})$. A
simple approximate expression, valid for $0\leq Z_{q}\leq 10$, is
obtained by writing $\delta_{q}(Z_{q})$ in the leading order of
magnitude
\begin{eqnarray}\label{82}
{{Z_{d}}^{1/2}\left( \tilde{m}_{s}-\tilde{m}_{d}\right)
\over \sqrt { \left( 1+\tilde{m}_{s}\right)
\left(  1-\tilde{m}_{d}\right)+2Z_{d} \left(
  \tilde{m}_{s}-\tilde{m}_{d} 
\right)}}-{{Z_{u}}^{1/2}
\left( \tilde{m}_{c}-\tilde{m}_{u}\right)
\over \sqrt { \left( 1+\tilde{m}_{c}\right)
\left(  1-\tilde{m}_{u}\right)+2Z_{u} 
\left( \tilde{m}_{c}-\tilde{m}_{u} \right)}}\simeq0.040~.
\end{eqnarray}
In this way, to each value of $Z_{u}$ corresponds one value of
$Z_{d}$. Since $Z_{u}$ is still a free parameter, to avoid
ambiguities, we may further assume that the up and down mass matrices
are generated following the same symmetry breaking pattern, that is,
\begin{eqnarray}\label{84}
Z_{u}=Z_{d} \equiv Z
\end{eqnarray}
Conditions (\ref{64}) and (\ref{84}) fix the value of $Z$.  

The numerical computation of $Z$ was made using the exact numerical
solutions of eq.~(\ref{42}). We found,
\begin{eqnarray}\label{86}
Z = {5\over 2}~.
\end{eqnarray}
The corresponding values of $(\delta_{u},\delta_{d})$ are
\begin{eqnarray}\label{88}
\delta_{u}=0.000064\qquad\qquad\qquad\delta_{d}=0.002300~~.
\end{eqnarray}

A discussion of the meaning of $Z_q \approx \frac{5}{2}~$ in terms of
the symmetry breaking pattern\newline $S(3)_L\bigotimes S(3)_R \supset
S(2)_L\bigotimes S(2)_R$ and explicit expressions for the
corresponding mass matrices (textures) will be published elsewhere
\cite{15}.

\section{Computation of $V_{CKM}$}

Once the value of $Z$ is fixed at $5/2$, the theoretical expression
for the entries in ${V^{th}}_{CKM}$, obtained from the exact
expressions for $O_{u}$ and $O_{d}$, are written in terms of the four
mass ratios
$(\tilde{m}_{u},\tilde{m}_{c},\tilde{m}_{d},\tilde{m}_{s})$ and only
one free parameter, namely the CP violating phase $\alpha$.  

We kept the mass ratios fixed at the values given in (\ref{60}) and
(\ref{62}) and made a new $\chi^{2}$ fit of $|{V^{th}}_{CKM}|$ to the
experimental values $|{V^{exp}}_{CKM}|$. The best value of $\alpha$
was found to be
\begin{eqnarray}\label{90}
\alpha=76.77^{\circ},
\end{eqnarray}
with $\chi^{2}=0.28$.
The corresponding best value of $|{V^{th}}_{CKM}|$ is
\begin{eqnarray}\label{92}
|{V^{th}}_{CKM}|=\pmatrix{
0.97531 & 0.22081 & 0.00254 \cr
0.22063 & 0.97457 & 0.03913 \cr
0.00928 & 0.03810 & 0.999231 \cr
},
\end{eqnarray}
which is to be compared with
\begin{eqnarray}\label{94}
\mid {V^{exp}}_{CKM} \mid=\pmatrix{
0.9745-0.9760 & 0.217-0.224 & 0.0018-0.0045 \cr
0.217-0.224 & 0.9737-0.9753 & 0.036-0.042 \cr
0.004-0.013 & 0.035-0.042 & 09991-0.9994 \cr
},
\end{eqnarray}
we see that the agreement between computed and experimental values of
all entries in $|V_{CKM}|$ is very good.  For the Jarlskog \cite{14}
invariant we found the value
\begin{eqnarray}\label{96}
J=-2.128\times10^{-5},
\end{eqnarray}
in good agreement with current data on CP violation in the 
$K^{\circ}-\bar{K}^{\circ}$ mixing system.

\section{Concluding remarks}

In this work we made a detailed comparison of the experimentally
determined ${V^{exp}}_{CKM}$ mixing matrix with a theoretical
${V^{th}}_{CKM}$ derived from quark mass matrices obtained from a
simple scheme for breaking the flavour permutational symmetry. The
entries in ${V^{th}}_{CKM}$ are, in principle, functions of the four
mass ratios $m_{u}/m_{t},m_{c}/m_{t},m_{d}/m_{b}$ and $m_{s}/m_{b}$,
one CP violating phase, and two small parameters $\delta_{u}$ and
$\delta_{d}$. To avoid a continuous ambiguity in the numerical fitting
procedure, we impose a phenomenologically motivated constraining
condition on the possible values of $\delta_{u}$ and $\delta_{d}$.  By
further assuming that the symmetry breaking pattern is the same in
both, $u$- and $d$-sectors, we fix the numerical value of the ratio
$Z=(M_{23}/M_{22})^{2}$ at $5/2$. In this way, we arrive at an
expression for ${V^{th}}_{CKM}$ written in terms of the four mass
ratios $m_{u}/m_{t},m_{c}/m_{t},m_{d}/m_{b}$ and $m_{s}/m_{b}$, and
only one free parameter, namely, the CP violating phase $\alpha$. A
$\chi^{2}$ fit of the matrix of the absolute values
$\mid{V^{th}}_{CKM}\mid$ to the experimentally determined
$\mid{V^{exp}}_{CKM}\mid$ gives the best value of
$\alpha=76.7^{\circ}$ and the value $J^{th}=-2.128\times10^{-5}$ for
the Jarlskog invariant, in good agreement with experiment. The
agreement between $\mid{V^{th}}_{CKM}\mid$ and
$\mid{V^{exp}}_{CKM}\mid$ is also very good with $\chi^{2}=0.28$.

\section{Acknowledgements}

One of us, E. R-J is indebted to Dr. J. R. Soto for help in the
numerical calculations. This work was partially supported by
DGAPA-UNAM under contract No. PAPIIT-IN110296.

\end{document}